# Correlating synthesis, structure and thermal stability of CuBi nanowires for spintronic applications by electron microscopy and *in situ* scattering methods

*Alejandra Guedeja-Marrón,*[1,2] *Henrik Lyder Andersen,*[3] *Gabriel Sánchez-Santolino,*[1,2] *Lunjie Zeng,*[4] *Alok Ranjan,*[4] *Inés García-Manuz,*[5,6] *François Fauth,*[7] *Catherine Dejoie,*[8] *Eva Olsson,*[4] *Paolo Perna,*[5] *Maria Varela,*[1,2] *Lucas Pérez*[1,5] *and Matilde Saura-Múzquiz*[1*]

[1]Departamento de Física de Materiales, Facultad de Ciencias Físicas, Universidad Complutense de Madrid, Madrid 28040, Spain

[2]Instituto Pluridisciplinar, Universidad Complutense de Madrid, Madrid 28040, Spain

[3]Instituto de Ciencia de Materiales de Madrid (ICMM), CSIC, Madrid 28049, Spain

[4]Department of Physics, Chalmers University of Technology, Gothenburg 41296, Sweden

[5]IMDEA Nanociencia, Madrid 28059, Spain

[6]Departamento de Física de la Materia Condensada (IFIMAC), Universidad Autónoma de Madrid, 28049 Madrid, Spain

[7]CELLS-ALBA Synchrotron, Barcelona 08290, Spain

[8]European Synchrotron Radiation Facility (ESRF), Grenoble 38000, France

*Corresponding author: matsaura@ucm.es

## Abstract

Bi-doped copper ($Cu_{1-x}Bi_x$) nanowires (NWs), promising candidates for spintronic applications due to their potential for a giant spin Hall effect (SHE), were synthesized and their structural properties and thermal stability were investigated. Using template-assisted electrodeposition, $Cu_{1-x}Bi_x$ nanowires with varying bismuth (Bi) content ($x$=0, 2, 4, and 7%) and different crystalline domain sizes were fabricated. Structural analysis by advanced electron microscopy and X-ray scattering techniques revealed the influence of synthesis conditions on the resulting NW crystal structure and microstructure, including Bi localization (within the lattice or in the grain boundaries), crystallite domain dimensions, and lattice distortions. While NWs with larger crystalline domains allow homogeneous Bi incorporation into the Cu lattice, NWs with smaller crystalline domains exhibit noticeable Bi accumulation at grain boundaries. The thermal stability of the NWs was examined using variable temperature X-ray diffraction and total scattering. Upon heating, lattice distortions consistent with Bi diffusion out of the Cu lattice were observed, with subsequent crystallization of rhombohedral metallic Bi upon cooling. These findings establish a foundation for optimizing the SHE performance of $Cu_{1-x}Bi_x$ nanowires for spintronic devices by correlating synthesis parameters with microstructural features and thermal behavior.

# Introduction

Bismuth (Bi) possesses one of the largest spin-orbit interactions (SOI) among atoms. Consequently, Bi has been incorporated into semiconductor heterostructures to leverage the strong SOIs induced by Bi for efficient spin control of electrons in semiconductor channels.[1,2] In the case of metals, significant effort has been made to utilize the strong SOI for producing spin-charge interconversion (SCI),[3] with the aim of incorporating Bi layers in spintronics devices. Surprisingly, in most experiments to date, the SCI efficiency measured in Bi is smaller than that in Pt, Ta, or W, which have smaller SOI than Bi. Only recently, large SCI has been measured in Ni/Bi(110) structures, ascribing the lack of previous results to the large anisotropy of the effective g-factor of Bi.[4] The effect is different when introducing Bi atoms as dopants in a metallic matrix. In particular, Bi impurities in a Cu host were identified by *ab initio* calculations as best candidates for all-metallic spin-current generation.[5] This calculation was experimentally verified by Niimi *et al.* who reported a substantial spin Hall angle (SHA) of approximately -0.24 in CuBi alloys with ~0.5% of Bi doping.[6] The presence of spin Hall effect (SHE) in highly Bi-doped Cu films has been confirmed by direct interface-free X-ray spectroscopy measurements in highly Bi-doped Cu films.[7] Although the mechanism of skew scattering has been proposed as the main driving force of the observed extrinsic spin Hall effect in CuBi alloys,[8] either the formation of extremely small clusters or the influence of interface roughness and grain boundaries decorated with Bi atoms may also be responsible for the observed phenomenon.[9] Other effects linked to the large SCI of Bi atoms in a Cu matrix have been recently reported, like a spin mixing conductance in CuBi/YIG larger than the one shown by similar Pt/YIG structures,[10] or the possibility of having large spin-orbit torque efficiency,[11] reflecting the potentiality of these new alloys for developing spintronics devices. However, before incorporating them into devices, it is crucial to have a deeper understanding of the mechanism behind the SOI-related effects and their correlation with the microstructure of the CuBi alloys. To achieve this, the alloys must be prepared using a synthesis/growth method that allows systematic control of sample characteristics such as composition, crystal quality, microstructure, and cluster formation.[12] Electrochemical deposition is a highly effective technique to produce nanowires (NWs) enabling precise control over both their composition and crystallinity in the case of CuBi alloys.[13]

To elucidate the relationship between grain size and spin transport properties, it is essential to examine systems with crystallite-sizes both smaller and larger than the reported spin diffusion length, estimated to range from 300 to 500 nm depending on the temperature.[14-17] For instance, NWs with smaller grain sizes (<500 nm) provide an opportunity to analyze how grain boundaries act as potential scattering sites for spin currents, most likely reducing SHE efficiency. Conversely, NWs with larger grains (>500 nm) allow spin transport properties in a regime with minimized grain boundary effects to be examined, highlighting the intrinsic properties of the CuBi alloy. Furthermore, understanding and controlling the thermal stability of these NWs is particularly critical for addressing fundamental questions about the performance of these materials under applied current, where temperature increase due to Joule heating is expected. Thus, unveiling the interplay between the synthesis method, micro/crystal/local structure, Bi doping, as well as potential structural changes that may arise in the system due to heat generation, is crucial for the rational optimization of this class of materials.

In the present work, we examine the structural characteristics of $Cu_{1-x}Bi_x$ NWs with different crystallite sizes and Bi contents prepared by electrodeposition. Given its established role in generating a significant SHA, we systematically analyze $Cu_{1-x}Bi_x$ NWs with Bi concentrations of 0%, 2%, 4%, and 7% to explore the impact of this heavy atom on the material's structure. The structural analysis is conducted over multiple length scales—atomic, nano, and micro—using a combination of advanced characterization techniques, including scanning transmission electron microscopy (STEM) and electron energy-loss spectroscopy (EELS), complemented by variable temperature synchrotron powder X-ray diffraction (SPXRD) with Rietveld analysis and X-ray total scattering (TS) with pair distribution function (PDF) analysis. We demonstrate how varying Bi concentration and electrodeposition synthesis conditions leads to changes the microstructure, such as Bi distribution within the NWs, crystallite size and lattice distortions. Furthermore, we conduct an in-depth investigation of the structural response of these systems upon heating using *in situ* variable temperature scattering experiments. While *ex situ* techniques often fail to capture the true behavior of materials under real-world operating conditions, *in situ* methods can provide real-time insights into *e.g.* phase transitions, atomic diffusion, and strain-induced effects taking place when subjected to real world operating conditions. The ability to fabricate $Cu_{1-x}Bi_x$ NWs with tailored structural characteristics and stability opens new avenues for designing materials with optimized SHE. By understanding the structure and thermal stability of high-quality electrodeposited $Cu_{1-x}Bi_x$ NWs, we establish a foundation for the precise fabrication of NWs with enhanced SHE and optimized spin transport properties for the development of next-generation spintronic devices.

## Results and discussion

**Morphology and microstructure of NWs - STEM/4D STEM**

Bi-doped Cu NWs with about 50 nm in diameter and varying amounts of Bi content ($Cu_{1-x}Bi_x$ with $x$=0, 0.02, 0.04, 0.07) were synthesized by template-assisted electrodeposition within the pores of anodized aluminum oxide (AAO) templates as described in the methods section.[18] Notably, the length of the synthesized NWs is determined by the growth time, while the diameter of the NWs is determined by the pore diameter of the AAO template resultant from the anodization process. As we have previously reported,[13] the growth of polycrystalline NWs with different crystalline domain sizes can be achieved by tuning the overpotential during electrodeposition. It has been shown that faster deposition rates generally lead to smaller crystallites due to enhanced nucleation [19], whereas slower rates favor growth of larger crystallites.[20] Reducing the overpotential during the growth, leads to a lower current density, producing a more uniform and consistent electrodeposition, which results in improved crystallinity.[21] However, employing low current densities may also affect the conductivity of the electrolyte,[22] and can prevent deposition from taking place. In order to lower the electrodeposition rate without reducing the conductivity of the electrolyte, an alternative approach is needed. Chelating agents, such as acids, are often used to enhance the solubility of metallic ions, due to their capability to form complexes with metals.[23] We believe the formation of such metallic complexes may also be a handle to tune the growth rate by slightly "hindering" the electrodeposition of the metals and therefore reducing the crystallization speed. If effective, different concentration of chelating agents could lead to different crystallite sizes in the obtained NWs. In this study, tartaric acid (TA) was used as chelating agent [24]. Hence, during the synthesis process, TA was incorporated into the electrolyte to improve

Bi solubility, allowing a better tuning of the crystallite sizes in the $Cu_{1-x}Bi_x$ NWs.[25] Electrodeposition growth curves showed that NWs synthesized with lower TA concentrations exhibit a growth time of 7500 s, whereas those with higher TA concentrations require 12000 s for growth. However, despite the change in electrodeposition rate due to TA, the current densities were in the same order of magnitude in both cases, suggesting that the same amount of Bi is being incorporated into the NWs in both cases, but at a slower electrodeposition rate in the case of high TA concentration. These results indicate that the use of TA does effectively reduce the electrodeposition rate.

A total of eight $Cu_{1-x}Bi_x$ samples were prepared using different concentrations of $Bi(NO_3)_3$ and TA. An overview of the synthesized samples is given in Table 1. SC/LC refer to small crystallites (SC) or large crystallites (LC), and the number describes the percentage of Bi in the resulting $Cu_{1-x}Bi_x$ NWs as determined by STEM-EELS (described later). Thus, samples of three different nominal concentrations were prepared for both the SC and LC type of NWs. These are referred to as SC2, SC4, SC7, LC2, LC4 and LC7 respectively, indicating whether they form NWs with small or large crystallites, and the approximate mean Bi content (2%, 4% and 7%, respectively). Two additional reference samples of pure Cu and pure Bi NWs were also synthetized.

Table 1. Overview of synthesized samples including sample name, concentration of $Bi(NO_3)_3$ and TA, and electrodeposition potential. Samples with an asterisk, with a more pronounced difference in TA

concentration, were used for 4D-STEM measurements.

| Sample name | Bi(NO$_3$)$_3$ concentration (mM) | TA concentration (M) | Electrodeposition potential (V) |
|---|---|---|---|
| SC2 | 2 | 0.33 | -0.05 |
| SC4 | 4 | 0.33 | -0.05 |
| SC7 | 8 | 0.33 | -0.05 |
| SC7* | 8 | 0.33 | -0.05 |
| LC2 | 2 | 0.99 | -0.05 |
| LC4 | 4 | 0.99 | -0.05 |
| LC7 | 8 | 0.99 | -0.05 |
| LC7* | 8 | 1.32 | -0.05 |
| Cu | NA | 0.33 | -0.3 |
| Bi | 8 | 0.33 | -0.07 |

Figure 1(a, b) shows representative low magnification STEM images of NWs released from the alumina template for the SC7* and SL7* samples, which were synthesized with high and low concentration of TA, respectively (see Table 1 for sample details). The samples consist of NWs with average lengths of around 10 μm as the cross-sectional SEM view of the NWs embedded in the AAO template depicts (see Figure S2). However, in the STEM images some NWs were found to be shorter, likely due to local fractures.

## HAADF-STEM

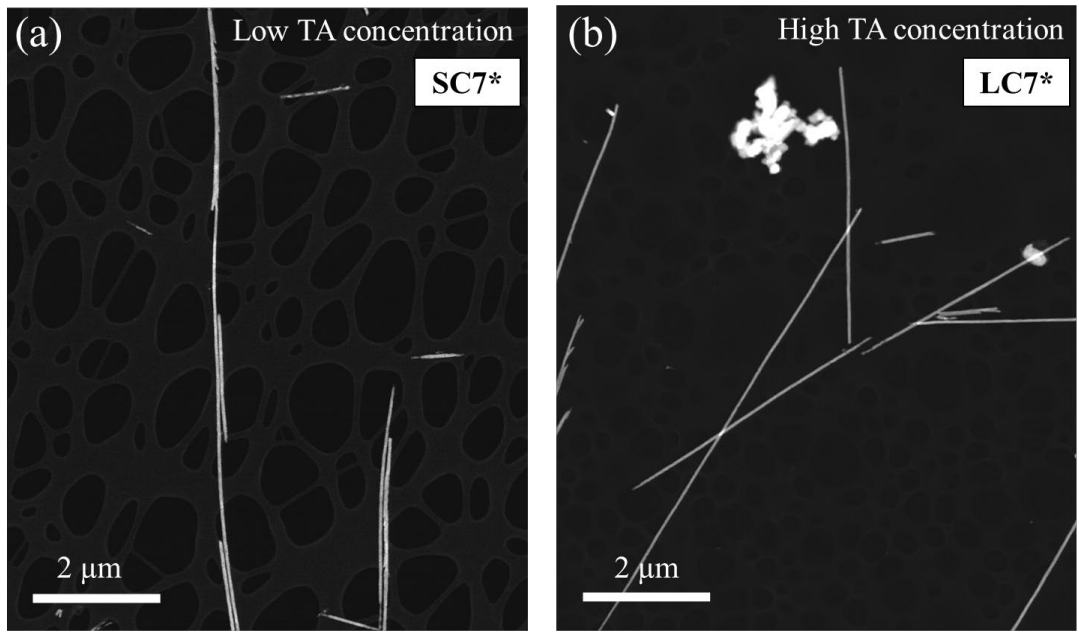

## 4D-STEM

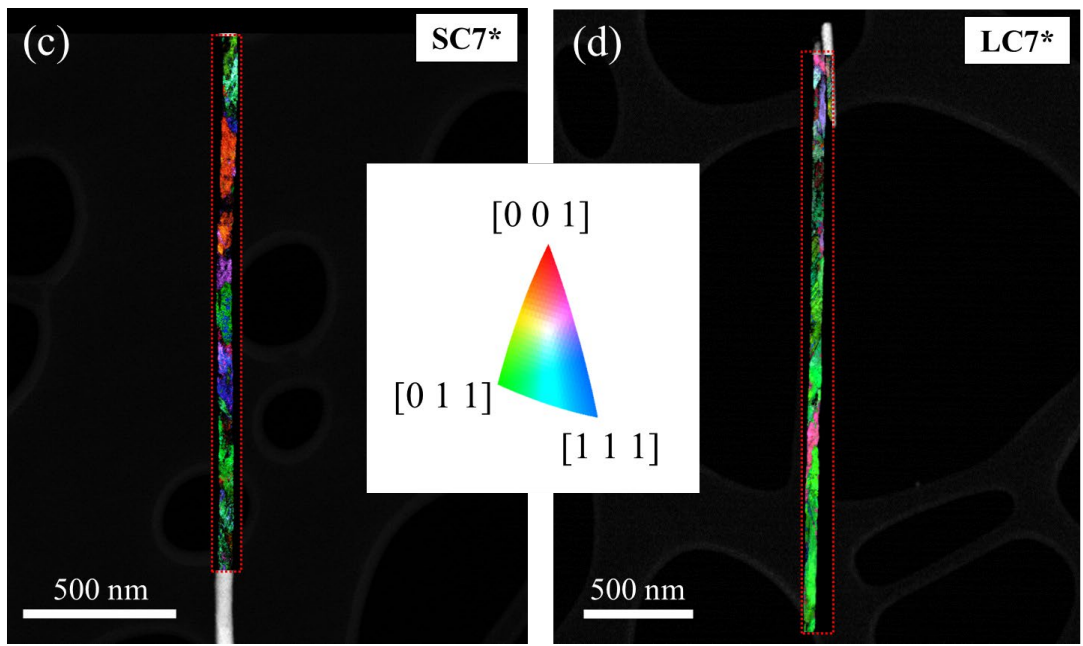

Figure 1: Top panel: High-angle annular dark field (HAADF) Low magnification STEM images of Bi-doped Cu NWs synthesized with (a) low concentration (SC7*) and (b) high concentration (LC7*) of TA. Bottom panels: Crystal orientation maps derived from 4D-STEM datasets of (a) SC7* NW and (b) LC7* NW showing crystalline domains and out-of-plane crystalline orientation colored by the stereographic projection of the [001], [011] and [111] directions.

To investigate whether changing the TA concentration during the synthesis has an influence on the resulting sizes of the crystalline domains of the $Cu_{1-x}Bi_x$ NWs, four-dimensional STEM (4D-STEM) spatially resolved nanodiffraction measurements were performed to obtain crystal-orientation maps.[26] Figure 1(c, d) shows the crystal orientation maps of two NWs from samples SC7* and LC7*, grown from electrolytes with different TA concentration: SC7* (0.33 M TA), LC7* (1.32 M TA) (see Table 1). The maps show a clear difference in the crystallite sizes of both samples. The SC7* NW (lower TA concentration) shows smaller crystallites of approx. 200 nm, whereas SC7* (higher TA concentration) shows much larger crystallites of around 1 μm. Thus, the samples synthesized with low concentration of TA are referred to as small crystallite-size (SC) nanowires, and those synthesized using higher TA concentration are referred to as large crystallite-size (LC) nanowires.

**Composition and Bi distribution - EELS/EDS**

To determine the nominal Bi content within the grains of the different $Cu_{1-x}Bi_x$ NWs, EELS measurements were conducted on NWs grown with varying $Bi(NO_3)_3$ concentrations in the precursor electrolyte (see Table 1). Figure 2 illustrates the observed spatial distribution of the relative Bi atomic percent within the grains for three NWs samples in the LC series (high TA concentration) grown using different $Bi(NO_3)_3$ concentration in the electrolyte. The Bi distribution was mapped using EELS, and the relative Bi content (in atomic percent) maps for each NW were obtained using the Gatan Digital Micrograph standard routines. Figures 2(a, b, c) display the HAADF images of the areas sampled for samples LC2, LC4 and LC7, respectively. White rectangles mark the region where EEL spectrum imaging was performed on each specimen. Cu $L_{2,3}$ and Bi $M_{4,5}$ absorption edges were studied in order to extract relative Bi/Cu compositions. Figure 2 (d-f)) displays the resulting Bi relative compositional maps for each material, again LC2, LC4 and LC7 from right to left, respectively. Each panel exhibits two maps that actually correspond to the same data. The Bi relative composition appears quite inhomogeneous, so contrasts have been manually tuned to highlight visually the spatial compositions within the inner core (marked with bule rectangles) and the outer shell (highlighted with green rectangles) regions for the sake of visual clarity. Histograms for each dataset have been constructed for statistical quantification, shown in Figure 2 (g-i) again same order of samples from left to right. Two main contributions are present in all maps, associated with the core and the native oxide layer on the surface of the NWs, formed due to air exposure.[27-29] To distinguish the two contributions, the histograms were fitted using the superposition of two Gaussian curves, highlighted in blue (core) and green (shell). In all cases a significant Bi segregation in the form of few nm thick Bi oxide on the shells is detected (O map shown in what follows), where the local Bi concentration doubles with respect to the core. But all cores exhibit a relatively homogeneous Bi distribution within the grains, with a direct correlation between the concentration of $Bi(NO_3)_3$ in the electrolyte, and the resulting Bi atomic percentage in the NWs, with average Bi compositions of 1.4%, 4.9% and 7.6% respectively for the three samples (with a 1% uncertainty in all cases). Since the three different concentrations of $Bi(NO_3)_3$ that were tested (2, 4, 8 mM), should yield approximate Bi doping within the core of the NWs of 2%, 4% and 7%, respectively, the EELS mapping gives proof of accurate control of the composition of $Cu_{1-x}Bi_x$ NWs.

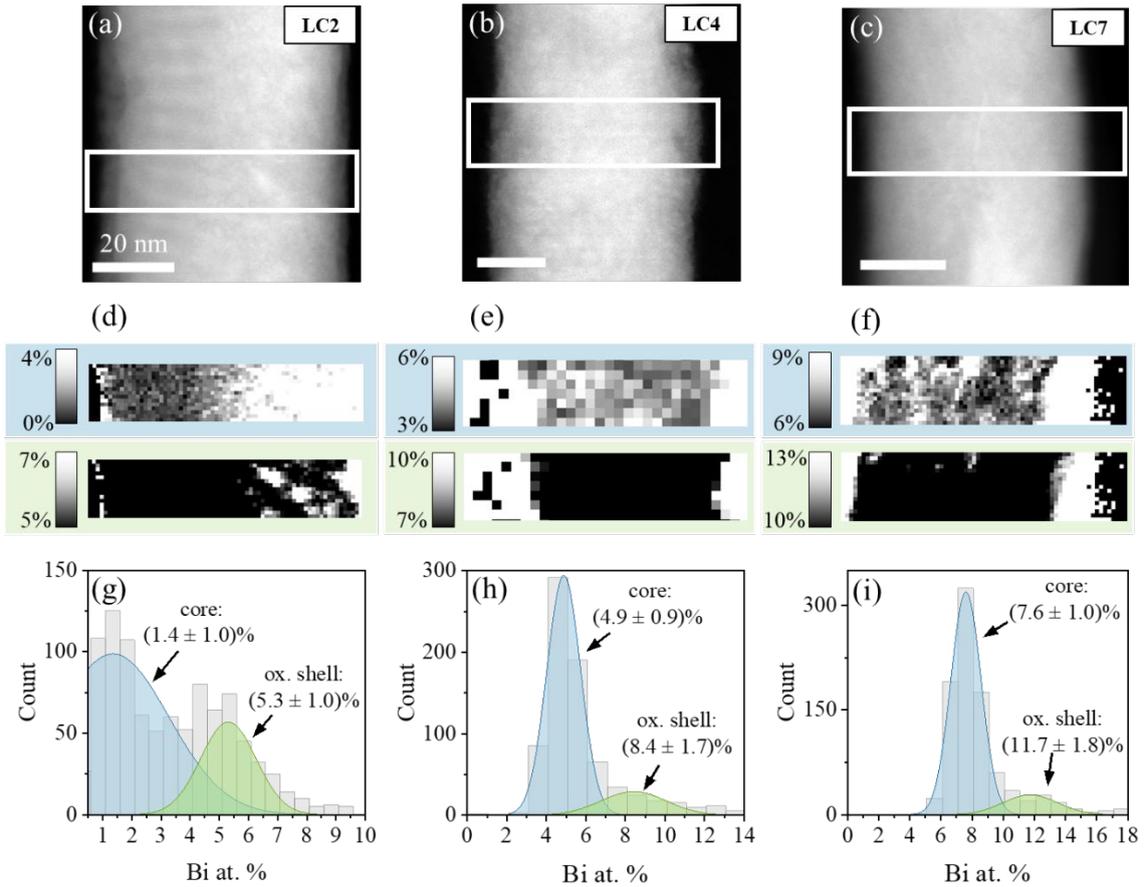

Figure 2: (a-c) ADF-STEM images of a single NW from samples (a) LC2, (b) LC4 and (c) LC7. (d-f) Spatially resolved maps of the relative Bi composition (in atomic %) within the regions marked with white rectangles in a-c, respectively. Top (blue rectangle) and bottom (green rectangle) maps in each panel show the same data, with different contrasts adjusted for visual purposes to highlight the contribution of core (blue) and oxide shell (green) regions. (g-i) Corresponding histograms quantifying the Bi content resulting from data in (d) LC2, (e) LC4 and (f) LC7 NWs. The two Gaussian curves fitted to the EELS histograms correspond to the Bi contributions from the core (blue) and shell (green) regions.

For comparison between the SC and LC series, EELS maps were also collected on both SC7 and LC7 NW samples. Figure 3 displays low magnification HAADF images (top panels) of SC7 (a) and LC7 (b) NWs, along with maps showing the integrated signal under the Bi $M_{4,5}$ edge in false color scale extracted from EEL spectrum images, measured within the regions highlighted with white rectangles for both samples (bottom panels).

These elemental maps indicate that Bi is not distributed equally within the grains of the two types of NWs. For the SC7 sample Bi accumulates preferentially in the grain boundaries. Meanwhile, for the LC7 samples, where larger grains are formed and less grain boundaries are present, Bi is distributed homogenously throughout the grains. These results indicate that tuning the synthesis method, not only provides a handle for controlling the crystallite sizes of the NWs, but it also leads to significant differences in the distribution of the Bi dopant within them. This difference in Bi distribution may lead to variations in the spin length diffusion of the NWs, which in turn would influence the SHE response.[8]

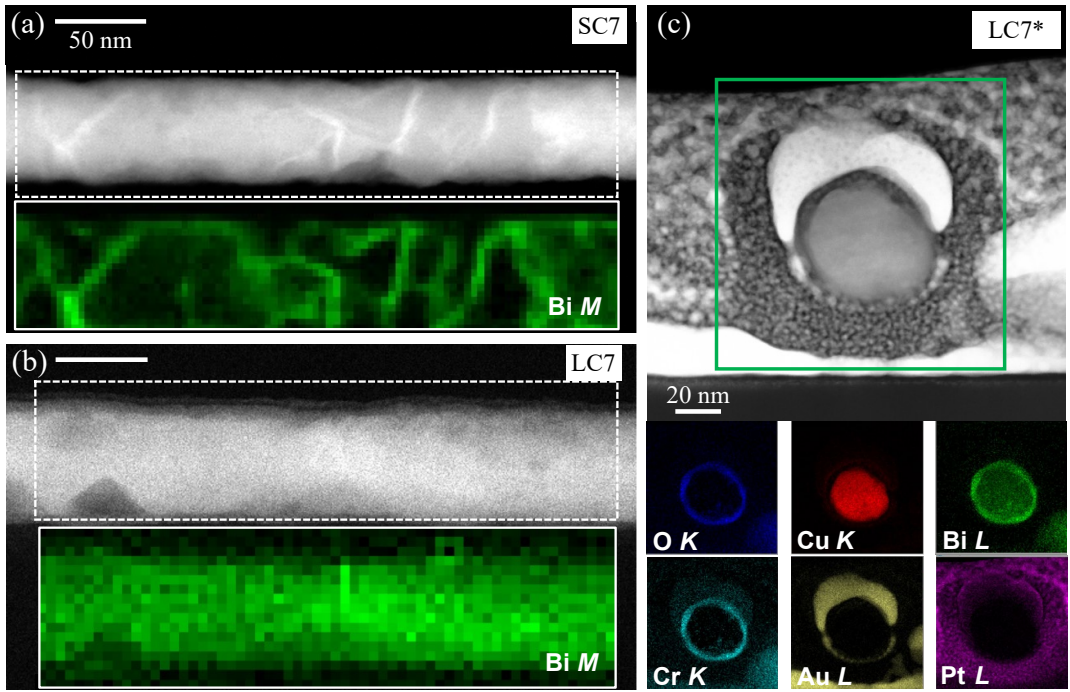

Figure 3: (a-b) ADF-STEM low magnification images of (a) SC7 NWs and (b) LC7 NWs, with inserts below showing the Bi $M_{4,5}$ edge EELS signal (green) extracted from the white squared regions. (c) HAADF image of the cross-section view of a (LC7*) NW and corresponding EDS elemental maps derived from the O K, Cu L, Bi L, Cr K, Au L and Pt L edges from top to bottom and left to right, respectively.

To further investigate the spatial homogeneity of the Bi radial distribution within the LC

NWs, a cross-section of a few LC7* NWs dispersed onto a Si substrate was prepared by FIB-SEM (see Figure 3c and Figure S3). After drop-casting the NWs on a Si substrate and before the FIB-SEM process, the NWs were coated with a layer of Au using magnetron sputtering to prevent ion damage. In Figure 3(c), an ADF image of the cross-section of one of the LC7* NWs is shown. The NW core along with several shell regions are clearly defined. The bottom panels depict a series of elemental maps obtained by EDS, including all O $K$, Cu $L$, Bi $L$, Cr $K$, Au L and Pt $L$ edges of interest. The Cu core is well defined, with a significant Bi signal that is homogeneously distributed in the radial direction. The fact that both the Bi and Cu signals are homogeneously distributed throughout the NW core, verifying the effective insertion of Bi within the Cu lattice, as well as the lack of discrete Bi clusters confirms again the high quality wire growth. A Bi oxide shell is clearly detected as well, with some Cr contamination present as well. Such surface layers of Cr and Bi oxides are a few nm thick at most, and result of the chemicals ($H_2CrO_4$ solution) used during the release of the NWs from the nanoporous AAO template and the oxidation due to exposure to air. The Au and Pt protecting layers derived from the conventional lamella preparation by FIB-SEM method are also appreciated around the NWs.

**Crystal structure – Room temperature (RT) SPXRD**

To investigate the average crystal structure of the samples and whether secondary oxide phases or metallic Bi clusters are formed in the as-synthesized NWs (still embedded in the AAO template), high-angular resolution SPXRD data were collected for SC and LC samples with different Bi doping (see Table 1). Figure 4 shows the Rietveld refinement of SPXRD data collected at room temperature (RT) on NWs from the two series (SC and LC) with different Bi doping levels (0%, 2%, 4% and 7%). For reference, and given that different unit cell parameters are often obtained for nanosized and bulk structures of the same compounds, a sample of pure Cu NWs (0% doping) and a sample of pure Bi NWs were also grown using the same method as for the Bi-doped Cu samples. The refined RT lattice parameter of our synthesised Cu NWs is 3.61446(1) Å, which is slightly smaller than that reported for bulk, defect free Cu of 3.61491 Å at 25 °C.[30] It is worth noting that metallic Cu and Bi do not crystallize in the same structure. While Cu crystallizes in the cubic *Fm*-3*m* space group, metallic Bi crystallizes in rhombohedral *R-3m* structure. Illustrations of the Cu and Bi crystal structures are shown in Figure S4, and the Rietveld refinement of SPXRD data collected on the pure Bi NWs sample is given in Figure S5. In the refinements, the atomic positions and occupancies were kept fixed for both phases, while scale factors, unit cell parameters, zero shift and an overall isotropic thermal parameter ($B_{iso}$) were refined. The peak profiles were modelled using the Thompson–Cox–Hastings formulation of the pseudo-Voigt function, using a platelet-vector-size model.[31] Notably, the instrumental contribution to the total peak broadening was determined by refinement of data collected on a NIST LaB$_6$ 660b calibrant in the same instrumental configuration and deconvoluted from the sample broadening in the refinements. Since the Cu structure contains just a single Wyckoff site, the amount of incorporated Bi within the lattice cannot be extracted from the SPXRD data, as the occupancies of different elements (although they have different scattering factors) on the site is fully correlated with the scale factor.[32] Therefore, all $Cu_{1-x}Bi_x$ phases were refined as pure Cu. Although this may lead to a slight error in the refined weight fractions of Bi-rich, Bi-poor and metallic Bi phases (see discussion later), the effect is minor due to the small percentage of Bi present in the samples, leading to a potential error of approx. 2% in refined weight fractions (see Table S1). Notably, the peak shapes of the secondary nanosized metallic Bi phase were fitted assuming spherical strain-free crystallites.

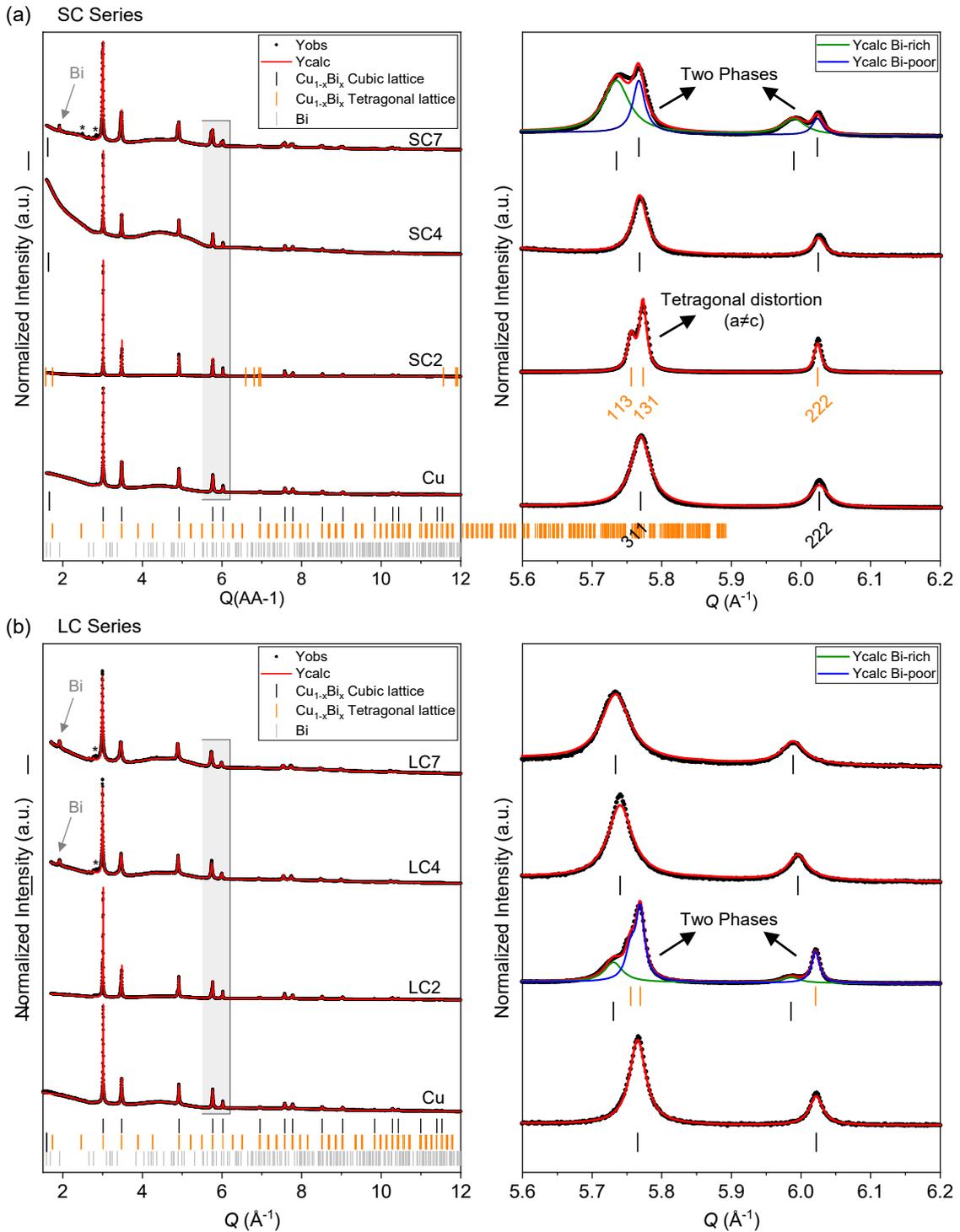

Figure 4: Room temperature powder X-ray diffraction patterns and corresponding Rietveld fits for the (a) SC and (b) LC sample series of nanowires with different Bi doping concentrations (0%, 2%, 4%, 7%). The positions of the Bragg peaks of the cubic and tetragonal $Cu_{1-x}Bi_x$ structures are given by black and orange bars, respectively, and in gray bars for rhombohedral Bi. When two phases are present, the calculated model

for each phase is given by a green and blue line on the right hand-side panels. The Miller indices of the cubic and tetragonal phases shown in the right hand-side panel of a) are given to illustrate the peak splitting taking place in the tetragonal distorted lattice. Minor peaks corresponding to unknown impurities (Q~2.9 Å$^{-1}$) are marked with an asterisk.

The high-angular resolution SPXRD data revealed that, in the SC series (Figure 4(a)), the sample with 2% Bi (SC2) exhibits a tetragonal distortion of the lattice, with $a=b=3.61001(1)$ Å and $c=3.62468(1)$ Å. This is clearly observed by the splitting of e.g. the 200 reflection (at Q~3.48 Å$^{-1}$) into two distinct reflections (002 and 020), and the 311 reflection (at Q~5.77 Å$^{-1}$) into two distinct reflections (113 and 131) as observed in the enhanced Q-regions on the right in Figure 4(a). This tetragonal distortion is not observed for the SC4 or SC7 samples, which are both cubic. However, a consistent splitting of all diffraction peaks is observed for the sample with the highest Bi content (SC7). This consistent splitting does not correspond to a tetragonal distortion, as observed for SC2. Rather, in SC7, the cubic *Fm*-3*m* lattice splits into two cubic phases of different unit cell volumes. The presence of the two cubic phases suggests the formation of two $Cu_{1-x}Bi_x$ alloys of different composition, one being Bi-poor and another one being Bi-rich. The larger sublattice, most likely being Bi-rich (considering the atomic radii of Cu=1.35 Å vs. Bi=1.60 Å),[33] constitutes the majority of the sample (70.4(5) wt%), and it exhibits a larger broadening of the peaks, indicating smaller crystalline domains and/or higher strain of the Bi-rich phase compared to those of the Bi-poor phase. Notably, the Bi-rich phase remains cubic rather than exhibiting the rhombohedral structure of metallic Bi. In addition, a small amount of metallic Bi in rhombohedral *R*-3*m* (1.75(8) wt%) was also observed in the SC7 sample (see Figure S6), indicating that some metallic Bi clusters also formed during the electrodeposition process.

Figure 4(b) shows the Rietveld refinement of the SPXRD data from the samples in the LC series. Here, the higher Bi-content NWs (LC4 and LC6) consist of a single homogenous cubic phase, while the LC2 sample shows the presence of two cubic crystalline phases with different unit cell volumes (similar to the observation for SC7). Furthermore, careful inspection of the data reveals that the major $Cu_{1-x}Bi_x$ phase in the LC2, which constitutes 66.9(4) wt%, also exhibits a tetragonal distortion of the lattice equivalent to the one observed in the SC2 sample, with a unit cell of $a=b=$3.61108(1) Å and $c=$ 3.62295(3) Å. The minority phase (33.1(4)%) was refined as cubic with $a=b=c=$3.63640(4) Å. The refined weight fractions, unit cell parameters and unit cell volumes of all samples are given in Table 2.

Table 2: Refined weight fractions, lattice parameters and unit cell volumes of all phases in all samples. For SC7 and LC7, where two $Cu_{1-x}Bi_x$ phases are present, in addition to the refined parameters for each phase, the calculated weighted average (WA) parameters of both $Cu_{1-x}Bi_x$ phases are given.

| Sample | Phase | wt.% | $a$-axis (Å) | $c$-axis (Å) | UC volume (Å$^3$) |
|---|---|---|---|---|---|
| **Cu** | Cu | 100.0(2) | 3.61446(1) | - | 47.2205(4) |
| **Bi** | Bi | 100.0(4) | 4.54464(3) | 11.8661(1) | 212.245(3) |
| **SC2** | $Cu_{1-x}Bi_x$ | 100.0(3) | 3.610010(7) | 3.62468(1) | 47.2375(2) |
| **SC4** | $Cu_{1-x}Bi_x$ | 100.0(4) | 3.615389(9) | - | 47.2565(4) |
| **SC7** | $Cu_{1-x}Bi_x$ (Bi-poor) | 27.8(3) | 3.61612(2) | - | 47.2856(7) |
| | $Cu_{1-x}Bi_x$ (Bi-rich) | 70.4(5) | 3.63650(3) | - | 48.090(1) |
| | WA $Cu_{1-x}Bi_x$ | 98.3(3) | 3.63073(2) | - | 47.8619(9) |
| | Bi | 1.75(8) | 4.5440(5) | 11.874(3) | 212.32(6) |
| **LC2** | $Cu_{1-x}Bi_x$ (Bi-poor) | 66.9(4) | 3.61108(1) | 3.62295(3) | 47.2429(5) |
| | $Cu_{1-x}Bi_x$ (Bi-rich) | 33.1(4) | 3.63640(4) | - | 48.086(2) |
| | WA $Cu_{1-x}Bi_x$ | 100.0(3) | 3.61946(2) | 3.62740(2) | 47.5208(5) |
| **LC4** | $Cu_{1-x}Bi_x$ | 93.2(5) | 3.63036(2) | - | 47.8464(8) |
| | Bi | 6.8(2) | 4.5194(8) | 11.955(4) | 211.46(9) |
| **LC7** | $Cu_{1-x}Bi_x$ | 93.8(6) | 3.63448(2) | - | 48.010(1) |
| | Bi | 6.2(3) | 4.519(1) | 11.921(6) | 210.9(1) |

The refined unit cell volume of samples from both the SC and LC series is plotted in Figure 5 as function of Bi content obtained from the EELS data. For samples where two phases are present (LC2 and SC7), both of the refined unit cell volumes are shown in green and blue for the Bi-rich and Bi-poor phases respectively, as well as the weighted average unit cell volume in black. In those cases, the weight percentages of the two cubic phases are also given. As Figure 5 indicates, the LC series exhibits a gradual, almost linear increase in average unit cell volume with increasing Bi content (considering the weighted average unit cell volume in LC2). This is consistent with Vegard's law,[34] and confirms effective Bi incorporation into the Cu lattice for the entire sample (not only in isolated NWs). In contrast, the SC series shows minimal lattice expansion with increasing Bi content for the SC2 and SC4 samples, and for the minority phase (Bi-poor, blue) of SC7. The main phase of SC7 (Bi-rich, green), however, exhibits a unit cell volume closer to that of the LC7 sample.

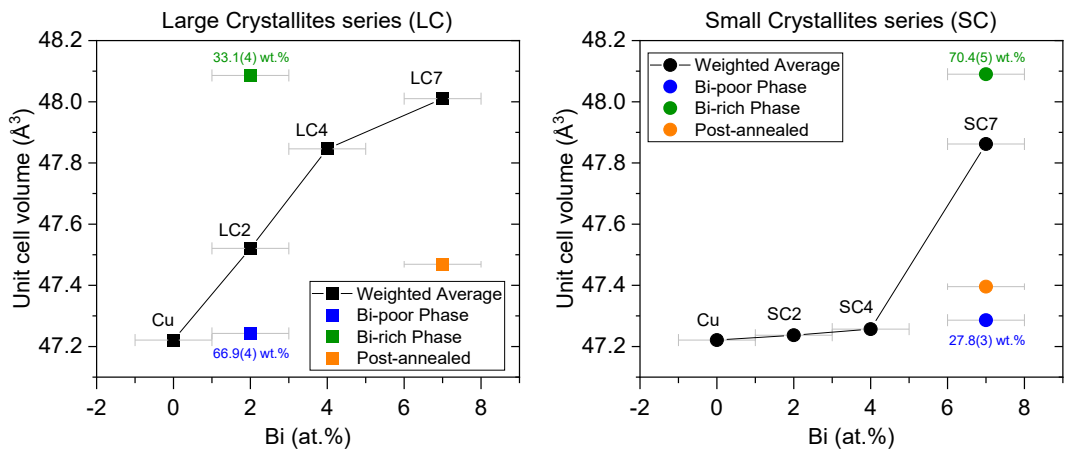

Figure 5: Refined unit cell volumes as a function of Bi content (extracted from EELS data) for the large crystallite (LC) and small crystallite (SC) series. For LC2 and SC7, where two Cu phases are refined, the unit cell volume of both phases is given in green and blue, with the black symbol corresponding to the weighted average unit cell volume. The errors on the refined unit cell volumes (error on Y axis) are smaller than symbol sizes.

The subtle volume change in the SC2 and SC4 samples corroborates that only a negligible amount of Bi is effectively incorporated into the crystalline lattice and, instead, most of the Bi localizes at the grain boundaries as observed in the EELS data (see Figure 3). The same trend is also followed by the Bi-poor phase of SC7. However, the main phase of SC7 (70.4(5) wt.%) does exhibit the expected lattice expansion within the grains due to Bi insertion. This suggests that for sufficiently high Bi concentrations, incorporation of Bi within the grain also takes place for SC NWs. The Bi at the grain boundaries is highly disordered or amorphous and therefore does not give rise to Bragg reflections in the SPXRD data. These results indicate that for low Bi content, the accumulation of Bi in the SC series appears in a non-crystalline, boundary-localized form, contributing little to the average lattice distortion within grains, whereas in the LC samples, homogenous insertion of Bi into the crystalline lattice leads to a consistent increase in lattice volume with Bi content, both for the low and high Bi content samples. This demonstrates that the STEM+EELS observations are representative of the entire sample and thus indicate that, through the reported synthesis/growth method, it is possible to control the Bi content and its distribution, although further investigation is needed to determine how the crystalline and amorphous Bi distribution influences the SHE of the material.

Notably, the crystallite sizes of the NWs cannot be accurately determined from SPXRD data in this system. The 4D STEM data reveal crystallites of several hundred nanometers in size along the length of the NWs, which is above the resolution limit for X-ray diffraction size determination. While the crystallite diameter falls within the resolution range (approximately 50 nm), extraction of anisotropic size information is unreliable due to the cubic symmetry of the lattice. This symmetry results in equivalent lattice planes along both the nanowire length and diameter, rendering impossible the extraction of trustworthy crystallite sizes from the Rietveld refinements.

**Thermal stability – Variable temperature (VT) SPXRD with slow heating and cooling**

Understanding the thermal stability of the studied NWs is essential, as their potential use in spintronic devices involves exposure to electric currents, which can induce significant temperature increases due to Joule heating. To examine the thermal stability of the $Cu_{1-x}Bi_x$ NW structure, variable temperature SPXRD data were collected on the SC7 sample upon heating to 450°C and subsequent cooling (see Figure 6(a) and Figure S7) A heating ramp of 2°C/min was applied from RT to 450 °C (~3.5 hours) followed by cooling at 20°C/min (~0.5 hour) back to room temperature, while sequentially collecting diffraction patterns with a time resolution of 1 minute. Sequential Rietveld analysis was carried out on the collected data, and the refined unit cell parameters and weight fractions are given in Figure 6(b-c). As discussed previously, two $Cu_{1-x}Bi_x$ phases as well as a minor metallic Bi phase were observed for the SC7 sample. As seen in Figure 6(a-b), when the heating starts (in the RT – 100 °C region), the unit cell parameters of both the Bi-rich (~70 wt%) and Bi-poor (~30 wt%) $Cu_{1-x}Bi_x$ phases increase linearly with the same slope due to thermal expansion. In the case of the Bi-poor phase, this linear expansion behavior continues up to a temperature of approx. 250 °C, at which this phase starts to disappear. Meanwhile, the Bi-rich phase follows an entirely different trend: At approximately 100 °C the increase in the unit cell slows down, reaching a maximum unit cell of 3.64116(4) Å at 143 °C. From that point, the unit cell exhibits negative thermal expansion, reducing in size with increasing temperature up to approx. 300 °C, above which it starts to increase linearly again.

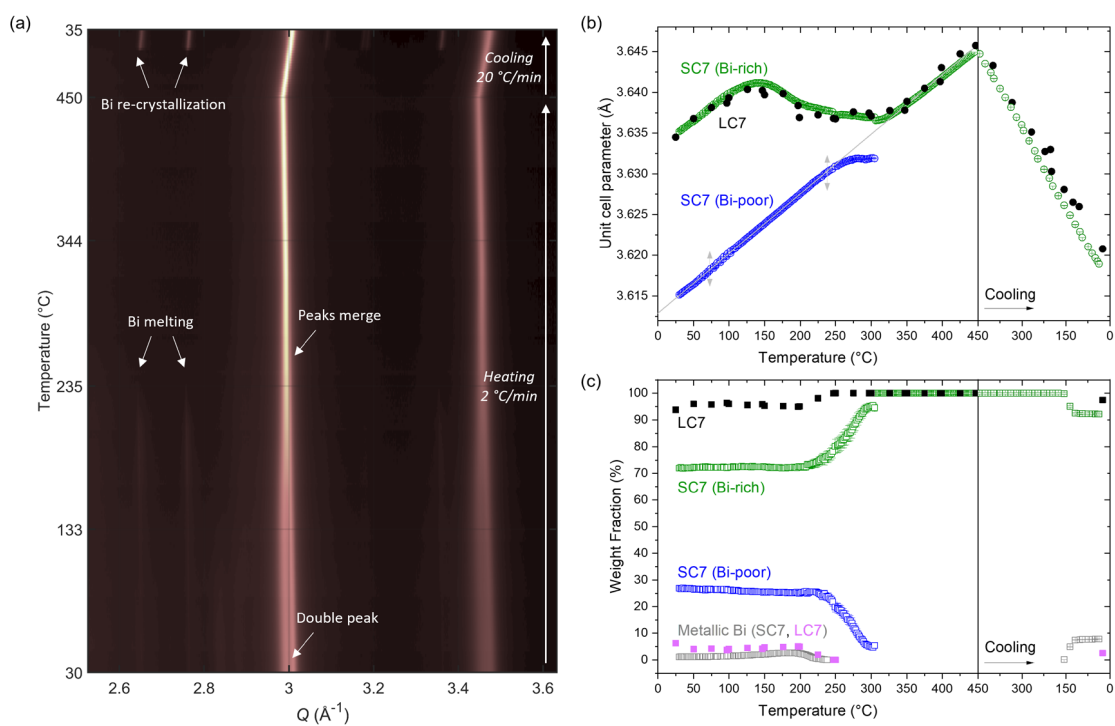

Figure 6. a) Contour plot of selected $Q$-region of VT SPXRD data collected on the SC7 sample. Full $Q$-range data of SC7 and LC7 samples can be found in Figure S7 and S8, respectively. b) Refined unit cell a-parameter of the $Cu_{1-x}Bi_x$ phase in LC7 (black), and of both $Cu_{1-x}Bi_x$ phases of SC7 (Bi-rich in green, and Bi-poor in blue) as function of temperature. The linear fit parameters of the SC7 Bi-poor initial thermal expansion (grey line) are given in Table S2. For the equivalent graph of LC7 showing the temperature range RT-1000 °C, see Figure S9. c) Refined weight fractions as function of temperature.

At the same time, as shown in Figure 6(c), the amount of metallic Bi, which initially accounts for 1.13(7) wt%, increases slightly reaching 2.63(7) wt% at 182 °C. These results indicate that, at approx. 100 °C, Bi starts diffusing out of the Cu lattice in the Bi-rich phase, causing the reduction in unit cell volume of the $Cu_{1-x}Bi_x$ phase and the increase in weight fraction of crystalline Bi in rhombohedral $R$-$3m$ space group. This takes place up to 300 °C, where no further diffusion of Bi from the $Cu_{1-x}Bi_x$ phase takes place, leading to the subsequent linear increase in unit cell parameter following thermal expansion. Notably, above 182 °C the crystalline Bi phase diffraction peaks start to gradually diminish, completely disappearing at approximately 250 °C, which corresponds to the melting of Bi (Bi melting point = 271.4 °C).[35]

Upon cooling back to RT, the lattice shrinks as expected with decreasing temperature. At approximately 150 °C, metallic Bi re-crystallizes, reaching 7.8(1) wt.% at RT. Once cooled to RT, the unit cell parameter of the single cubic $Cu_{1-x}Bi_x$ phase present is 3.618930(7) Å. This parameter is smaller than the initial unit cell of the Bi-rich phase (3.63650(3) Å), but larger than that of the Bi-poor phase (3.61612(2) Å). Moreover, it is 0.12% larger than the unit cell of the pure Cu NWs sample (3.61446(1) Å). This finding indicates that not all Bi has diffused out of the lattice, but a smaller amount remains present after thermal treatment, and this amount of Bi doping is stable within the probed temperature range.

An equivalent trend in unit cell size with increasing temperature is observed for the LC7 sample, where a single $Cu_{1-x}Bi_x$ phase was initially observed with a unit cell parameter similar to that of the Bi-rich SC7 (see black symbols in Figure 6(b-c)). Not only is the trend the same, but both samples exhibit practically identical values in unit cell parameter as function of temperature. This result confirms that the diffusion of Bi due to heat is consistent and it takes place at the same temperatures across different $Cu_{1-x}Bi_x$ samples. As was observed in the SC7 sample, the diffraction peaks of metallic Bi initially present in LC7 also disappear at approximately 250°C. The LC7 sample was heated up to 1000 °C, to test the thermal stability to higher temperatures (see Figure S8). The lattice parameter from 450 °C onwards follows a linear increase due to thermal expansion (see Figure S9). At 850 °C, the alumina template crystallizes into nanosized domains of γ-alumina (see Figure S8 and S10), which remains crystalline once cooled down to RT. Like for the SC7 sample, in LC7 a small amount of Bi (2.6(1) wt.%) recrystallizes when cooled to RT after being heated to 1000°C. However, it is a lower amount than what was initially present in the as-synthesized LC7 (6.2(3) wt.%). The lattice parameter of the Cu$_{1-x}$Bi$_x$ phase of LC7 once cooled down to RT is 3.62077(1) Å. This result, once again, corroborates that there is a certain amount of Bi that remains in the NWs after heating, and it is stable up to high temperatures of 1000 °C.

**Thermal stability – Variable temperature (VT) SPXRD with fast heating and quenching**

In order to further investigate thermal stability (specifically whether faster heating and cooling rates influence the diffusion of Bi from the Cu matrix) as well as the local NW structure (see next section), variable temperature total scattering data were collected on the SC7 NW sample. Here, the sequential collection of total scattering datasets started at ambient conditions with 78 second time resolution. While continuously collecting TS data, the capillary was then rapidly heated to 400 °C, by translating in the hot air blower pre-heated to the target temperature. After 20 minutes at 400 °C, the capillary was quenched back to room temperature by retracting the hot air blower. Due to the small sample volume, almost instantaneous heating and quenching were achieved. The target temperature of 400 °C was chosen based on the slow heating experiment, which showed a stable unit cell size at 400 °C, with a linear increase in size due to thermal expansion.

Figure 7(a) shows a contour plot of the time-resolved synchrotron X-ray TS data (selected $Q$-range) collected on a new batch of sample SC7, denoted here as SC7-B, before, during and after heating to 400 °C for 20 min. At all three stages, the main crystalline phase observed has a cubic *fcc* structure (space group *Fm-3m*), *i.e.* isostructural to Cu (orange arrows in Figure 7a). As for the original SC7 sample discussed in the RT section, the high angle diffraction peaks of the TS data hint that two Cu$_{1-x}$Bi$_x$ *fcc* phases are present in the sample (*i.e.* a Bi-rich and a Bi-poor phase). However, due to the high energy of the beam, the two phases cannot be resolved in the TS data as the peak splitting is only slightly visible in the low intensity high Q peaks. Therefore, the sample used for TS experiments, denoted as SC7-B, was modeled with only a single $Cu_{1-x}Bi_x$ phase in a cubic *fcc* structure. In addition to the main $Cu_{1-x}Bi_x$ phase, a minor amount of crystalline metallic Bi (space group *R-3m*) is initially observed to be present (blue arrows). Upon heating the sample to 400°C, the secondary Bi phase melts (Bi melting point is 271.4°C)[35] resulting in the disappearance of the Bi Bragg peaks and emergence of a broad diffuse scattering peak (green arrow). This is consistent with the slow heating VT results, which showed melting of the metallic Bi at around 250°C. After 20 minutes of heating at 400°C, the sample was quenched, leading to recrystallization of a larger amount of Bi than initially present prior to heating, as evident from the emergence of more intense Bi peaks.

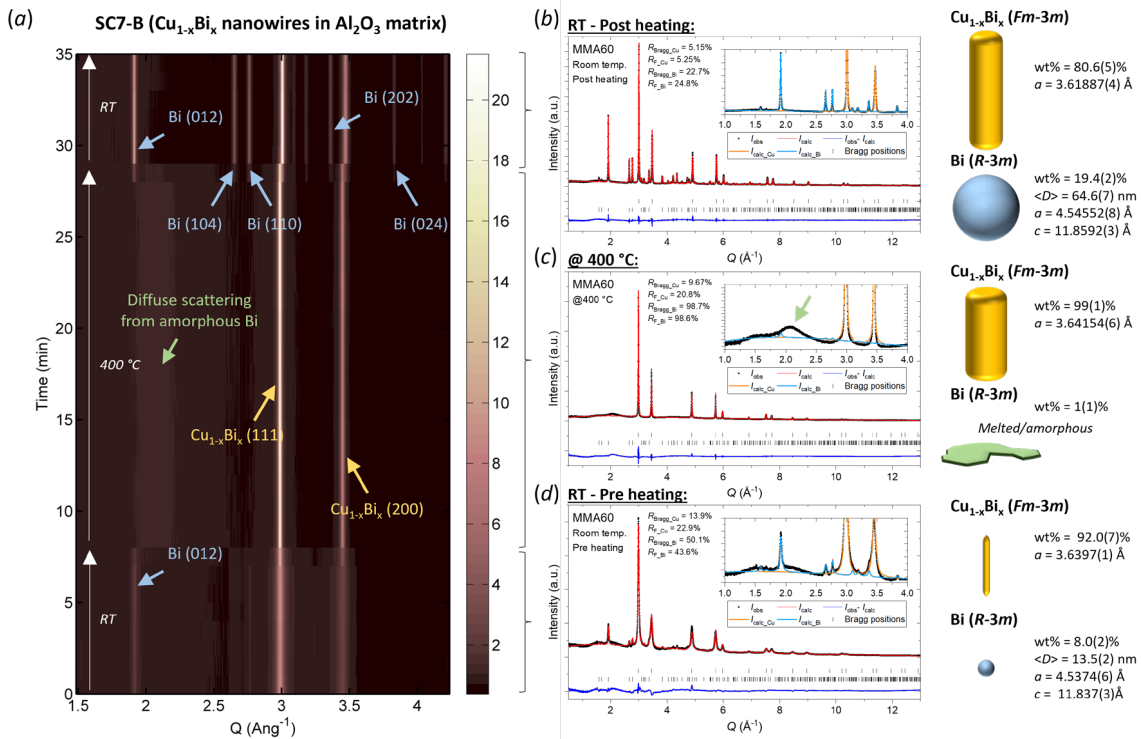

Figure 7: a) Contour plot of low *Q*-range of time-resolved X-ray TS data collected on the SC7-B $Cu_{1-x}Bi_x$ nanowire sample before, during and after heating to 400 °C. b-d) Rietveld analyses of the Bragg diffraction

in summed TS data from the three stages. The inserts illustrate the disappearance and recrystallization of the secondary Bi phase during the experiment. Selected compositional, structural and microstructural parameters from the refinements are provided on the right side of the corresponding data along with illustrations of the different phases.

Rietveld analysis was carried out for Bragg scattering in the summed X-ray TS datasets from the three stages (pre, during, post heating) to extract quantitative compositional, structural and microstructural information (see Figure 7(b-d)). Prior to heating, the main cubic $Cu_{1-x}Bi_x$ phase observed has a unit cell parameter of 3.6397(1) Å. The corresponding Rietveld refinement of the pure Cu NWs sample (see Supporting Information) yielded a lattice parameter of 3.61416(8) Å, which is consistent with the one obtained from Rietveld refinements of SPXRD collected at MSPD, ALBA for the same sample (3.61446(1) Å). Consequently, the larger cell parameter of the $Cu_{1-x}Bi_x$ sample can be concluded to be consistent with successful doping of the Cu matrix by the larger Bi atoms. In fact, the unit cell parameter of the $Cu_{1-x}Bi_x$ phase in SC7-B is slightly larger than the unit cell parameter obtained for the Bi-rich phase of the equivalent SC7 sample discussed in previous sections (3.6365(3) Å). This result is expected given that a single $Cu_{1-x}Bi_x$ phase is observed here, rather than having some of the Bi in a minor Bi-poor phase, as observed for the other sample.

As mentioned earlier, the high symmetry of the cubic $Cu_{1-x}Bi_x$ structure, which has only one Wyckoff atomic position, prevents refinement of the respective Cu and Bi occupancies, as they fully correlate with the scale factor causing fit divergence. Thus, the composition of the main cubic phase must be either inferred/estimated from the unit cell parameter or characterized by complementary techniques. In this case, in the multiphase refinements of all datasets, the atomic structure of the main phase was fixed to the nominal $Cu_{0.93}Bi_{0.07}$ composition established by EELS, although the observations indicate that the amount of Bi dopant within the Cu lattice is likely to decrease during/after heating. This was done to avoid overestimation of the $Cu_{1-x}Bi_x$ phase weight fractions (due to the higher X-ray scattering power of Bi compared to Cu), which, in turn, would lead to underestimation of the secondary crystalline Bi weight fractions. Thus, in addition to the 7 mol% of Bi in the nanowire lattice, a secondary Bi phase weight fraction of 8.0(2) wt% (equivalent to 3.0(1) mol%) was obtained prior to heating. Upon heating, the Bragg peaks of the Bi phase disappear while the lattice parameter of the $Cu_{0.93}Bi_{0.07}$ phase increases to 3.64154(6) Å. Whether the increase in lattice parameter is solely due to thermal expansions or may have a contribution from incorporation of some of the melted Bi into the structure is not clear from the TS data. However, given the obtained results from the slow heating experiments, where heat leads to extraction of Bi from the Cu lattice, it is safe to assume that the increase in lattice parameter at 400 °C is exclusively due to thermal expansion. While the scale factor parameter for the Bi phase was allowed to refine during the heating step, the resulting minor crystalline Bi weight fraction of 1(1) wt% is undoubtedly an artefact arising due to background correlations (see insert in Figure 7(c)). Upon quenching, a considerably larger amount of secondary Bi phase (*i.e.* 19.4(2) wt% or 7.8(9) mol%) than initially present prior to heating (*i.e.* 8.0(2) wt% or 3.0(1) mol%) is observed to recrystallize. This finding further supports that Bi was successfully incorporated into the original $Cu_{1-x}Bi_x$ nanowire structure, and that it has diffused outside the Cu lattice due to heating. Considering the additional crystalline Bi formed (7.8(9) mol% – 3.0(1) mol% = 4.8 mol%), this would indicate that approximately 2.2 mol% of the original 7 mol% Bi still remains in the Cu lattice. The refined post heating cell parameter of 3.61887(4) Å also remains slightly larger than the 3.61416(8) Å obtained for the pure Cu nanowires, which is consistent with some Bi remaining in the structure, and with the results obtained in the slow heating/cooling experiment (final unit cell parameter 3.618930(7) Å). Notably, equivalent measurements were conducted on released $Cu_{1-x}Bi_x$ nanowires, to check whether the alumina template influences the observed Bi diffusion. However, the same results were observed for the embedded and released NWs (see *Released $Cu_{1-x}Bi_x$ nanowires* in Supporting Information).

**Local atomic structure: Pair distribution function analysis.**

Determining the position of Bi in the structure by conventional powder X-ray diffraction (Bragg scattering) analysis is not possible unless the Bi is highly ordered across many unit cells giving rise to superstructure peaks, none of which were observed here in the high-angular resolution SPXRD data. Given the relatively low amount of Bi within the samples, even in the highly doped ones (7 at.% Bi), it is more likely to be incorporated in a disordered way or with only local order. In TS with PDF analysis, both the Bragg and diffuse scattering signal is utilized, allowing local structural features to be examined. While the nature of the Bi doping into the Cu lattice is most likely substitutional, *i.e.* occupying one of the Cu atomic positions in the lattice, it could in principle also be interstitial, *i.e.* occupying one of the voids in the Cu structure. Notably, given the larger atomic radius of Bi (van der Waals radius = 207 pm) compared to Cu (van der Waals radius = 140 pm), the introduction of Bi in the lattice should give rise to considerable local strain, and possibly the formation of Cu vacancies. The obtained PDF of the for the SC7-B sample prior to heating (see Figure S12) shows no peaks indicating the presence of interstitial Bi. Rather, the experimental PDF is in good agreement with the simulated PDF of the cubic Cu structure (see Supporting Information).

Figure 8 shows the PDFs and corresponding structural fits for the SC7-B sample before, during and after heating to 400°C. The real space structural fit to the PDF before heating is mostly in good agreement with the average crystal structure from the Rietveld analysis of SPXRD data discussed earlier. For the PDF prior to heating, the first two peaks are well fitted, while a misfit can be observed for the peaks at ~4.46, 5.75 and 6.81 Å (see black arrows). This finding could indicate that rather than Bi being substituted randomly into Cu positions in the lattice, some degree of local Bi ordering exists. At 400°C, the misfits discussed above disappear as the Bi either disorders or migrates out of the Cu lattice. Instead, a small feature appears at ~3.3 Å (green arrow), which may be associated with the predominant Bi-Bi bond length in the amorphous melted Bi at high temperature. Post heating, the fit to the main Cu phase remains good indicating that the majority of the Bi has either left the structure or remains disordered. Notably, a peak at ~3.1 Å associated with the metallic Bi appears (see Figure S12).

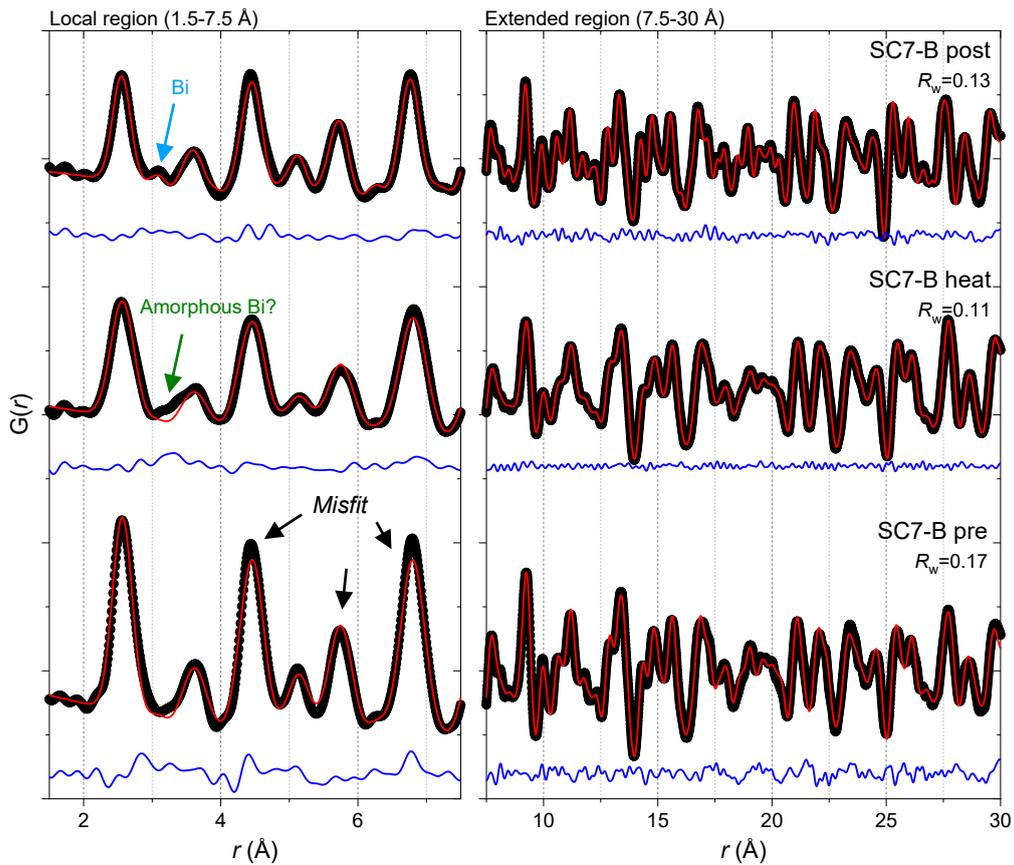

Figure 8: PDF fits for the SC7-B sample collected before (bottom), during (mid) and after (top) heating, all fitted by the same two-phase ($Cu_{0.93}O_{0.07}$ and Bi) model.

The observed misfits to the PDF of the $Cu_{0.93}O_{0.07}$ NWs prior to heating could indicate that the strongly scattering Bi atoms (compared to Cu) cannot sit next to each other in nearest neighbor ($r$=2.57 Å) or next-nearest neighboring ($r$=3.64 Å) sites, but rather preferably order on the sites with interatomic distance of ~4.46 and 6.81 Å. This idea is illustrated in the expanded (2x2x2) unit cell shown in Figure 9, where Bi has been placed on a set of specific positions separated by ~4.46 (black arrows) and 6.81 Å (green arrows). No further noteworthy misfits are observed for higher $r$-values indicating that the Bi ordering is limited to a length scale of <2 unit cells. Thus, to test the proposed Bi ordering, fits to the PDF of pre-heated SC7-B were carried out using both the disordered and ordered model (including metallic Bi as a secondary phase) limiting the fitting range to 10 Å. As shown in Figure 9, the quality of the fit to the local structure improves when using the ordered model, with the intensity of the peaks being more accurately described.

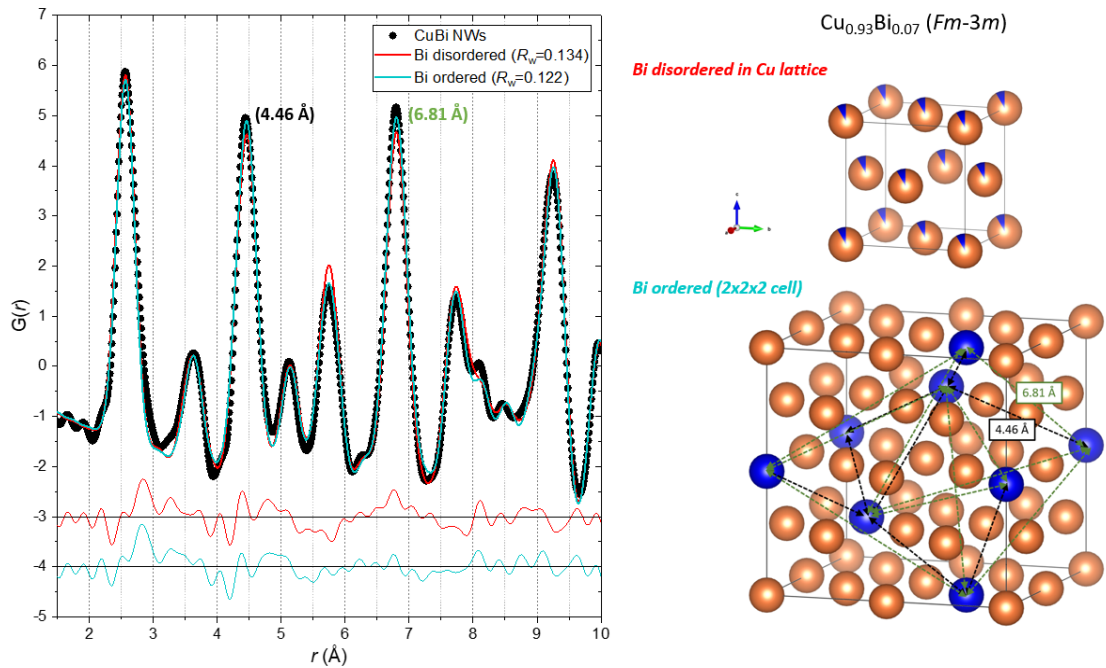

Figure 9: PDF of the pre-heated SC7-B sample fitted by structural models with disordered (red) and locally

ordered (cyan) Bi. The disordered and ordered structures are illustrated on the right. Atomic structures illustrated using the software VESTA.[36]

## Conclusions

This study provides a comprehensive investigation of the microstructure, structure and thermal stability of Bi-doped Cu nanowires synthesized using template-assisted electrodeposition. Macroscopically averaged X-ray diffraction techniques were combined with local electron microscopy measurements for this aim. The structural analysis reveals that, by varying the concentration of tartaric acid (TA) in the electrolyte solution during the template-assisted electrodeposition, control over crystalline domain sizes and Bi distribution is achieved. 4D-STEM analysis shows that smaller crystalline domains (~200 nm) are achieved at lower TA concentrations, resulting in Bi accumulation at the grain boundaries. In contrast, larger crystalline domains (~1 μm) obtained at higher TA concentrations promote uniform Bi incorporation into the Cu lattice, as evidenced by EEL spectra and a linear increase in refined lattice parameter with Bi doping, consistent with Vegard's law. To investigate the thermal stability of the NWs, variable temperature SPXRD data were collected to 450 and 1000°C, with slow heating (2 °C/min) and cooling (~20 °C/min) rates. Rietveld analysis of the data reveals that, above 100 °C, Bi begins diffusing out of the Cu lattice, with significant migration occurring between 150–350 °C. In addition, a small amount of metallic Bi (1-8 wt%) is often present in the as-synthesized samples, which melts at approximately 250 °C. When the samples are cooled down, metallic Bi (both initially present and diffused from the Cu matrix) recrystallizes. Importantly, both for slow and fast heating/cooling experiments, including those to 1000 °C, a fraction of Bi remains present in the Cu lattice once cooled back to RT, as evidenced by the values of the unit cell parameter of the post heated samples being consistently larger than that of pure Cu NWs. Results from all VT experiments indicate that the amount of Bi remaining in the lattice corresponds to approximately 1-2 % of Bi doping. PDF analysis of the TS data was carried out to examine the nature of the Bi integration in the Cu lattice. An improved fit to the PDF was obtained using a model with locally ordered Bi, which supports the hypothesis that Bi atoms in the Cu NWs exhibit a preference for specific interatomic distances rather than random distribution, with a tendency to avoid nearest and next-nearest neighboring positions. Our findings highlight the intricate relationship between synthesis parameters, structural characteristics, and thermal stability in this nanophase system. This study underscores the importance of tuning synthesis parameters to achieve targeted microstructural properties and thermal behavior, paving the way for the development of high-performance $Cu_{1-x}Bi_x$ nanowires tailored for next-generation spintronic devices. Elucidating the structural and thermal behavior of these materials is paramount for optimizing and tuning device performance under operational heating conditions. Nanowires with smaller crystalline domains, where Bi localizes at grain boundaries, may enhance or diminish spin scattering mechanisms critical for SHE. Further studies into the spin properties of the different NWs are needed to elucidate the structure-property correlation and thus understand and optimize the yet unknown mechanisms driving the giant SHE in CuBi alloys.

## Methods

**Synthesis of $Cu_{(1-x)}Bi_x$ nanowires**

$Cu_{1-x}Bi_x$ NWs ($x$=0, 0.02, 0.04, 0.07) were synthesized by template-assisted

electrochemical deposition in the pores of anodized aluminum oxide (AAO) templates with pore sizes of 50 nm and thicknesses around 40 µm, which were prepared by anodization onto ∅=25 mm of high-purity (99.999%) Al discs. The anodization was conducted for 5 hours in a two-electrode electrochemical cell at room temperature, using a 0.3 M oxalic acid ($C_2H_2O_4$) solution under continuous stirring. To maintain a constant temperature during anodization, the cell was placed on a copper plate equipped with a cooling circuit, through which a refrigerating liquid was circulated, connected to a cryothermostat (see Figure S1). The applied voltage was maintained at a constant 40 V with the aid of an adjustable DC power supply (EA Elektro-Automatik, Viersen, Germany). A Pt mesh was employed as counter electrode. Following anodization, the residual aluminum layer was chemically etched using a solution of 0.74 M $CuCl_2$ and 3.25 M HCl, while preserving an outer aluminum ring to provide mechanical support and facilitate handling in the following steps. Lastly, the pores were opened using a 5% (vol.) $H_3PO_4$ at room temperature for 1.5 hours. The complete anodization procedure has been described in detail in a previous publication from our group.[18] Prior to electrodeposition, one side of the nanoporous templates was coated with a Ti(15 nm)/Au(150 nm) thin layer deposited by sputtering, using a Leica EM ACE600 sputter coater. This conducting layer acts as a working electrode for electrodeposition.

The electrolyte was prepared by mixing varying amounts (0.14, 0.29 or 0.58 g) of $Bi(NO_3)_3·5\,H_2O$, 1.5 g of $CuSO_4·5\,H_2O$, 17.43 g of $KNO_3$ and varying amounts (7.44, 14.88, 22.32 g) of Tartaric acid (TA) ($C_4H_6O_6$) in 150 mL of a solution of 10% vol. glycerol in deionized water. This led to electrolyte solutions with final $Bi(NO_3)_3·5\,H_2O$ concentrations of 2, 4 and 8 mM, and different amounts of TA. The change in $Bi(NO_3)_3·5\,H_2O$ concentration was aimed at obtaining different compositions in the prepared $Cu_{1-}$ $_x$Bi$_x$ NWs, while the TA was added to examine its role as a chelating agent, specifically in enhancing Bi solubility and regulating crystallite sizes. HNO$_3$ was subsequently added to the mixture, under magnetic stirring, to lower the pH until complete dissolution of the electrolyte salts was achieved at around pH=0.9. The electrodeposition was carried out at room temperature using a three-electrode electrochemical cell with a Pt mesh as counter electrode and an Ag/AgCl (3M NaCl) electrode as reference electrode. The growth of the NWs was controlled using a Metrohm-Autolab PGSTAT potentiostat using an applied growth potential of -0.1 V during 10 s for nucleation and -0.05 V during the rest of the growth time. Note that both potentials were referred to the reference electrode. After the synthesis, the NWs were released from the AAO template using a 0.4 M H$_3$PO$_4$ and 0.2 M H$_2$CrO$_4$ solution, followed by repeated rinsing with milli-Q water. Once released, the NWs were stored in pure ethanol to avoid oxidation. Note that a vortex mixer rather than sonication was used for dissolution of the AAO template and subsequent rinsing to prevent the NWs from heating.

**Electron microscopy**

*STEM and EELS mapping*

STEM-EELS characterization was carried out using a JEOL JEM-ARM200cF aberration corrected electron microscope operated at 200 kV, equipped with a cold field emission gun and a Gatan Quantum spectrometer. For spectrum imaging, the electron beam was scanned along the region of interest, and EEL spectra were acquired with an energy dispersion of 1 eV/channel. Random noise was removed from the EELS data using principal-component analysis.[37] Background was subtracted using a power law fit before integration of the signal in order to produce EELS maps.

*Four-dimensional STEM nanodiffraction data collection and analysis for crystal-phase mapping*

4D-STEM nanodiffraction data were collected on a JEOL MonoNEOARM 200F instrument equipped with a monochromator, an imaging aberration corrector and a probe aberration corrector, a Gatan Imaging Filter Continuum HR electron energy-loss spectrometer (EELS) and an energy dispersive X-ray spectrometer (EDS). Data were acquired using a Gatan Rio Camera while operating at 200 keV with a convergence angle of 3.4 mrad and a camera length of 2.5 cm. The py4DSTEM package.[26] was used for the data calibrations and crystal-orientation mapping analysis. Calibrations include correcting shifts of the diffraction pattern, calibrating the rotational offset between the real and diffraction space, and calibrating the pixel sizes. After performing the calibrations, the detection of the Bragg peaks from each of the data points was carried out to obtain Bragg vector maps. Then, the average reciprocal lattice vectors were extracted and indexed.

*Lamella and EDS mapping*

LC7* NWs stored in ethanol solution were dispersed onto a Si substrate. To prevent them from ion damage during the lamella preparation by focus-ion beam (FIB), they were coated by sputtering with a 50 nm thick Au layer using a Leica sputter coater EM ACE600. Then FEI Versa 3D FIB-SEM was implemented to prepare the cross-sectional sample and the above mentioned JEOL MonoNEOARM 200F microscope was used for the lamella STEM imaging and EDS acquisition. The STEM observations were performed immediately after the preparation of the lamella via FIB in a coordinated manner to prevent oxidation of the NWs cross-section.

**Powder diffraction and total scattering experiments**

*Room temperature SPXRD*

The as-synthesized samples (amorphous AAO template containing the NWs) were crushed into a powder and packed into 0.3 mm Quartz capillaries, which were sealed using two-component epoxy glue. Room temperature (RT) high-angular resolution synchrotron powder X-ray diffraction (SPXRD) data were collected at the BL04-MSPD beamline at the ALBA Synchrotron with an energy of 30 keV,[38] and at beamline ID22 at the European Synchrotron Radiation Facility (ESRF),[39] with an energy of 35 keV.[40] The X-ray wavelengths were calibrated to 0.413560 Å (BL04-MSPD, 30 keV) and 0.354324 Å (ID22, 35 keV), respectively, from Rietveld refinement of powder X-ray diffraction data collected on a NIST $LaB_6$ 660b standard.[41] The capillaries were continuously spun during data collection to improve powder averaging. At BL04-MSPD, ALBA, the diffraction data were collected using the high-throughput position sensitive detector MYTHEN, while at ID22, ESRF, the SPXRD data were collected using the 13-channel Si 111 multi-analyzer stage and Dectris Eiger2 X 2M-W CdTe pixel detector.

*Variable temperature SPXRD*

Variable temperature (VT) SPXRD experiments with slow heating (2°C/min) and cooling (20°C/min) between RT and 450°C were also carried out at the BL04-MSPD beamline at the ALBA Synchrotron on selected samples, using the same conditions as for the RT measurements. The samples were heated using a FMB Oxford hot air blower (RT-950°C) mounted underneath the sample. The data were continuously collected with a one minute time-resolution.

*Variable temperature total scattering experiments*

Variable temperature X-ray total scattering (TS) experiments with instant heating to 400

°C, holding, and quenching to RT were conducted using the high-resolution powder diffractometer at beamline ID22 at the European Synchrotron Radiation Facility (ESRF).[40] The X-ray wavelength was calibrated to 0.206689(1) Å (60 keV) from Rietveld refinement of powder X-ray diffraction data collected on a NIST standard $LaB_6$ 660b standard.[41] Data were collected on the NWs still embedded in the alumina template (as for the RT measurements), as well as on released NWs. Both types of samples were packed into Quartz capillaries and sealed using two-component epoxy glue. Data were also collected under ambient conditions for an empty capillary and for a capillary containing an empty alumina template. During measurements, the capillaries were spun at 919 rpm and continuously rocked +/-1 mm to improve powder averaging. The samples were heated using a retractable Cyberstar hot air blower (RT-400°C) mounted underneath the sample. The hot air blower was preheated and translated in/out to achieve almost instantaneous heating/quenching of the samples. The scattering data were continuously collected with a 78 second time-resolution (sum of 25x 1 sec exposures + detector readout) using the Perkin Elmer 2D detector positioned at a distance of 380 mm behind the sample capillary giving a $Q_{max}$ of approx. 25 Å$^{-1}$.

*Structural analysis*

Rietveld analysis of the Bragg scattering in the diffraction and TS data was performed using the *FullProf Suite* software package.[42] Fourier transformation of the X-ray total-scattering (Bragg and diffuse sample scattering) data into real-space PDFs was carried out using *PDFgetX3*,[43] and the real-space structural refinements of the PDFs were carried out using the *PDFgui* software.[44] The background intensity from a 0.3 mm quartz capillary containing unfilled alumina template was subtracted from the TS data. The experimental $Q_{damp}$ (instrumental damping of PDF peak intensities) was determined to be 0.009 Å$^{-1}$ by fitting of NIST LaB$_6$ 660b calibrant PDFs collected in the same instrumental configuration.

## Acknowledgements

Financial support from the Spanish MCIN/AEI, Spain/10.13039/501100011033 (grants PID2021-122980OB-C51, PID2021-122980OB-C52, TED2021-131323B-I00, PID2020-117024GB-C43 and CNS2022-136143) and Comunidad de Madrid (TEC-2024/TEC-380 (Mag4TIC-CM), MAD2D-CM-UCM3 and MAD2D-CM-UAM) is acknowledged. A.G-M. and I.G-M. acknowledge support from MINECO through Grants RTI2018-097895-B-C43 and PRE2021-098702 respectively. M.S-M. gratefully acknowledges the funding from the European Union's Horizon Europe research and innovation programme under project No. 101109595 (MAGWIRE). H.L.A. gratefully acknowledges funding from the European Union's Horizon Europe research and innovation programme under project No 101063369 (OXYPOW). G.S.-S. acknowledges financial support from Grant RYC2022-038027-I funded by MICIU/AEI/10.13039/501100011033 and by ESF+. Authors acknowledge the node Centro Nacional de Microscopía Electronica (CNME) of ICTS "ELECMI" at Universidad Complutense de Madrid and Chalmers Material Analysis Laboratory (CMAL) at Chalmers University of Technology for the electron microscopy observations. The authors would also like to acknowledge ARTEMI to provide partial financial support for this work. We acknowledge the European Synchrotron Radiation Facility (ESRF) for provision of synchrotron radiation facilities under proposal number MA-5440 and CH-6713 and we would like to thank Andy Fitch for assistance and support in using beamline ID22. Experiments were also performed under proposal ID2022035793 at BL04-MSPD

## Associated content

**Supporting information**

The Supporting Information is available free of charge at XXXX

Electrodeposition setup. SEM micrograph of a cross section of an alumina template with grown $Cu_{1-x}Bi_x$ NWs. Additional HAADF-STEM images of a lamella cross-section view of several LC7* NWs. Representation of crystal structures. Additional Rietveld refinement fits of different samples. Comparison of refined weight fractions of the Bi-rich $Cu_{1-x}Bi_x$ phase refined as pure Cu *vs.* refined containing 7% Bi. Contour plots of variable temperature SPXRD data of SC7 and LC7 samples. Linear fit parameters of Figure 6(a). Refined unit cell parameters of LC7 as function of temperature. Contour plots, Rietveld refinements and additional information on released $Cu_{1-x}Bi_x$ NWs. Additional information on PDF analysis of SC7-B $Cu_{1-x}Bi_x$ nanowires.